# What is the speed limit of martensitic transformations?

Stefan Schwabe[a,b], Klara Lünser[a,b,c], Daniel Schmidt[d,e], Kornelius Nielsch[a,b], Peter Gaal[d,e], Sebastian Fähler[c,a*]

[a]*Leibniz IFW Dresden, Institute for Metallic Materials, Dresden, Germany;* [b]*TU Dresden, Institute of Materials Science, Dresden, Germany;* [c]*Helmholtz-Zentrum Dresden-Rossendorf, Institute of Ion Beam Physics and Materials Research, Dresden, Germany;* [d]*Leibniz-Institut für Kristallzüchtung (IKZ), Berlin, Germany;* [e]*TXproducts UG, Luruper Hauptstraße 1, Hamburg, Germany.*

*Corresponding author: Sebastian Fähler, Helmholtz-Zentrum Dresden-Rossendorf, Institute of Ion Beam Physics and Materials Research, Bautzner Landstraße 400, 01328 Dresden, Germany, s.faehler@hzdr.de

What is the speed limit of martensitic transformations?

Structural martensitic transformations enable various applications, which range from high stroke actuation and sensing to energy efficient magnetocaloric refrigeration and thermomagnetic energy harvesting. All these emerging applications benefit from a fast transformation, but up to now their speed limit has not been explored. Here, we demonstrate that a thermoelastic martensite to austenite transformation can be completed within ten nanoseconds. We heat epitaxial Ni-Mn-Ga films with a nanosecond laser pulse and use synchrotron diffraction to probe the influence of initial temperature and overheating on transformation rate and ratio. We demonstrate that an increase of thermal energy drives this transformation faster. Though the observed speed limit of 2.5 x $10^{27}$ $(Js)^{-1}$ per unit cell leaves plenty of room for a further acceleration of applications, our analysis reveals that the practical limit will be the energy required for switching. Thus, martensitic transformations obey similar speed limits as in microelectronics, expressed by the Margolus–Levitin theorem.

1. **Introduction**

"Can you make it faster?" is a question frequently heard by scientists. However, often physical limitations represent an insurmountable speed limit even for the best scientists. A famous example is microelectronics, where the clock frequency of computers already stopped its exponential increase two decades ago [1]. The physical limit in this case is the energy required for switching between two states, a speed limit predicted some time before it was reached [2]. Thus, it is worth to ask the question of the ultimate shortest time scale already when new functional materials and applications emerge. We focus on martensitic transformations, which switch the material's crystal symmetry within the solid state by temperature. These transformations are the underlying mechanism for the following emerging applications that will benefit from an increased speed since they use highly reversible cycles achievable only in so-called thermoelastic martensitic transitions. Actuation by shape memory effects is ultimately limited by the speed of this transformation [3]. The same limitation also holds for the reverse process used for sensing [4]. Martensitic transformations occur in the most promising magnetocaloric materials as well, where a higher switching frequency is required to increase the cooling power [5]. The reverse process, which is used to harvest low-grade waste heat by thermomagnetic materials, also benefits from an increased frequency [6]. Here, we experimentally approach the time limit required for a thermoelastic martensitic transformation – well before this speed limit will be reached in these applications.

What is a martensitic transformation and what is known about its speed? A martensitic transformation switches the crystal structure between high symmetry at high temperatures and low symmetry at low temperatures. It is a transformation without diffusion processes between a high temperature phase called "austenite" and a low temperature phase called "martensite". The atoms themselves move for much less than interatomic distances, which shears the crystal structure. This makes martensitic transformations a priori much faster than phase transitions that require the exchange of atoms by diffusion. The latter requires much higher temperatures and longer time scales, which limits the switching time e. g. of phase change memory technology [7]. Martensitic transformations are of first order and accordingly proceed by nucleation and growth, two processes, which occur at the microstructural level. Here we address the question, how fast a complete sample can transform in a fraction of a second, which comprises both underlying processes: the nucleation rate, and the growth velocity of these nuclei.

Steel, a prominent material showing a martensitic transition, has been quenched since historical times to increase its hardness, unwittingly benefitting from the martensitic phase. Nowadays, it is well known that in iron these irreversible transformations can propagate with a velocity of up to 1/3 of the speed of sound [8]. Further, early experiments revealed a shortest time scale of about 0.1 µs, as summarized by Nishiyama [9]. A similar time scale is observed by shock wave experiments [10]. Here, we focus on reversible martensitic transformations, as switching forth and back between both phases is required for the applications mentioned in the beginning. These so called thermoelastic transformations can be modelled by molecular dynamics [11,12], which indicates that a martensitic transformation can start and proceed already within several picoseconds. However, these calculations are limited to very short length scales and depend on boundary conditions. Experiments typically focus on the reverse transformation from the martensite back to the austenite during heating. This is because energy can be added much faster into a sample than heat can be removed by dissipation. Fast heating can for example be realized by passing a current through a shape memory wire [13]. By Joule heating, pulses as short as a µs have been realized and a subsequent transformation has been observed within 50…100 µs. However, for these experiments the wires had to be prestrained within the martensitic state in order to probe the change of length. The approach is invasive, as inertia limits the response time. Recently, this drawback has been overcome by probing stress instead of strain, which revealed that the transformation within the wires starts after around 20 µs [14]. Simultaneously performed in-situ diffraction experiments showed that the transformation at the wire surface starts as early as 10 µs [15]. In an alternative approach, a magnetic pulse of 13 ms was used to probe magnetic shape memory alloys [16]. Due to their magnetostructural coupling, this allows to induce a phase transformation, which was probed indirectly by a change of magnetization. In this case, dynamics are limited by the long pulse duration. Much shorter pulses can be realized by lasers, but this approach

5had been only applied for iron, which does not exhibit thermoelastic transformations. When a laser pulse was used to heat an iron film while recording microstructural changes by ultrafast electron microscopy, the bcc-fcc transformation occurred within 220 ns [17]. Finally, lasers with ultra high power can be used to induce shock waves, and the combination with dedicated in-situ diffraction experiments revealed that the stress induced fcc-hcp transformation in iron can occur within 2…4 ns [18]. In general, experiments are sparse especially for thermoelastic transformations as it is challenging to induce the phase transformation on short time scales and measure it non-invasively.

To probe how fast we can drive a thermoelastic martensitic transformation, we selected a dedicated setup (more details are given in the experimental section and supplementary Fig. S1). For heating, we used a 7 ns short laser pulse and chose a thin film sample with a thickness of 500 nm. Heat diffusion results in a temperature profile with acceptable homogeneity for this thickness on the time scale of the excitation pulse. As martensitic material, we selected Ni-Mn-Ga, a prototype for Heusler alloys, which exhibit excellent actuation [3,19], sensing [20] and caloric [5] properties. We used synchrotron X-ray diffraction to probe the structural transformation since this method measures the whole film thickness. The synchrotron provides X-ray pulses with a duration of 100 ps, which allows to measure dynamics in very short time scales. Moreover, a variable delay time between laser pulse and X-ray pulse makes our set-up ideally suited to probe the transformation in situ in the ns range. To obtain sufficient intensity for these pump-probe experiments, the films were grown epitaxially [21], which ensures that the diffracted intensity is concentrated in just a few peaks. In the examined film, the martensite to austenite transformation starts at around 336 K and is finished at around 365 K as obtained from diffraction measurements (see supplementary

Fig. S2 for details). To investigate the speed of the phase transformation and the amount of energy required, we varied the initial sample temperature and heating rate.

## 2. Materials and methods

The investigated, epitaxial sample was prepared by magnetron-sputter deposition in a UHV chamber (base pressure of around $2 \times 10^{-9}$ mbar) from a $Ni_{44}Mn_{32}Ga_{24}$-target. During deposition, the used MgO(0 0 1) substrate was heated to 400 °C and rotated to ensure a homogenous composition. For better adhesion of the thin film, a 20 nm Cr-buffer layer was deposited underneath the 500 nm $Ni_{50}Mn_{30}Ga_{20}$ film. The composition was measured with energy-dispersive X-ray spectroscopy (EDX) using an XFlash Detector on a JEOL JSM-6510 scanning electron microscope (SEM) with an accuracy of about 1 at.%. A SEM micrograph of the sample is shown as supplementary Figure S6.

For time-dependent diffraction measurements, the beamline P23 at DESY (storage ring PETRA III) was used. The beam energy was chosen to be 12.7 keV, which corresponds to a wavelength of 0.98 Å. To induce the martensite to austenite transition of the sample, the film was heated up with a nano-second laser (Ekspla NL202, pulse duration 7 ns (full width at half maximum, FWHM), wavelength = 1064 nm, repetition rate = 1 kHz). Though the laser light is absorbed by the electrons, on this time scale the electron and phonon systems are in equilibrium. The base temperature of the sample could be adjusted with a heater from Anton Paar. The laser energy was measured with a power meter for the different laser fluences used in the experiment. The laser spot of approximately 612 x 589 µm$^2$ exceeds the X-ray beam footprint of around 10 µm$^2$. Furthermore, the spot is much larger than the film thickness. Thus, we obtained structural information of a homogenously heated film fraction during the heating and subsequent cooling. The diffracted beam was detected with a X-spectrum Lambda 750

K detector, which has a spatial resolution of 55 µm and was positioned at a distance of around 0.34 m to the sample. A delay generator (Stanford Research DG645) ensured that the detector only summed up the intensity at particular times, which were selected depending on the different measurement routines. For the data points, the detected intensity for each time interval originated from a single bunch of electrons with an X-ray pulse duration of 100 ps. Due to some inherent characteristics of the laser, the delay at which the pulse is initiated changes depending on the used laser fluence. Accordingly, the time $t = 0$ was defined based on the gathered data points. It was set as the time when the intensity of the involved phases starts to show rapid changes due to the fast heating. All the other points in time are specified with respect to this zero-point.

## 3. Results

Diffraction is a non-invasive method to investigate a martensitic transformation by directly probing the changes of the crystalline structure. As an example, the diffractograms in Fig. 1 illustrate that heating with a 7 ns long laser pulse, which impinges the material at $t = 0$ ns, can be sufficient to fully transform the sample from martensite to austenite within 20 ns. Due to the required time resolution, the intensity in these diffractograms is low, despite examining a relatively thick film. In the limited time of our synchrotron experiment, we could only position our detector at two small key regions, which were selected following a recent crystallographic analysis of epitaxial films [22]. In the first region (Fig. 1(A-C)), the vicinity of $(0\ 0\ 4)_A$ austenite and $(\overline{14}\ \overline{2}\ 0)_{MM}$ modulated martensite peaks allows to observe both phases in a single detector image. In (A), the sample is martensitic before the laser pulse and is fully transformed to the austenite after around 20 ns in (C). At the end of the laser pulse at $t = 7$ ns (Fig. 1(B)), peaks of both phases are visible simultaneously and are separated, as expected for a first-order phase transition. For the second region, the $(\overline{16}\ \overline{2}\ 0)_{MM}$ peak of the martensite was selected, which is well isolated from all other peaks (Fig. 1(D-F)). The nearly complete transformation to the austenite is visible here as well, as the martensite peak disappears at $t = 20$ ns (Fig. 1(F)). For the following more detailed analysis of the time dependency, we used the summed-up intensity of this peak.

To understand the influence of fast heating, one first has to convert the applied laser fluence to a temperature rise $\Delta T^*$ with respect to the base temperature $T_0$ of the sample. As described in more detail within the supplementary (Fig. S3), we use the high brilliance of synchrotron radiation, which allows for a calibration of temperature through the thermal expansion of the austenite lattice [23,24]. We confirmed the validity of this conversion of laser fluence into $\Delta T^*$ by numerically calculating the temperature evolution from the thermal and optical properties of Ni-Mn-Ga (see supplementary Fig. S3). For the martensite to austenite transformation, the overheating $\Delta T$ is the most important quantity, which we define as the temperature increase above the austenite start temperature $T_{AS}$: $\Delta T = T_0 + \Delta T^* - T_{AS}$.

The influence of fast heating on the martensitic transformation is examined in two experimental series. In the first one, we varied the temperature rise $\Delta T^*$. To vary $\Delta T^*$ between 60 and 177 K, we increased the fluence of the laser pulse from 23 mJ cm$^{-2}$ to 60 mJ cm$^{-2}$ and recorded the diffraction patterns for different delay times. For each measurement, the base temperature $T_0$ was kept constant at around 329 K, which is just 7 K below the austenite start temperature and 36 K below the austenite finish temperature (see supplementary Fig. S2). Exemplary diffraction patterns around the $(\overline{16}\ \overline{2}\ 0)_{MM}$ peak before, during, and after the laser pulse are depicted in Fig. 1(D-F). For each time step, we summed up the peak intensity, which is proportional to the martensite fraction. From this, we directly obtain the time evolution of the martensite fraction (Fig. 2(A)), which proceeds similarly for all Δ$T$* examined. Before the laser pulse ($t < 0$ ns), the martensite intensity and thus fraction is high. As soon as the laser pulse hits the sample, the martensite fraction drops strongly within a few ns because the sample is heated and transforms into austenite. After the laser pulse, the martensite fraction increases again on a much longer time scale due to cooling. We first analyse the time interval where the amount of martensite phase decreases due to the laser heating and cover the time afterwards, when cooling takes place, in the next paragraph. When using a low laser fluence, which results in a relatively low Δ$T$* of 60 K, the martensite intensity decreases by about 30 % after the laser pulse. With increasing fluence, the fraction of martensite is reduced further, and the transformation proceeds faster. When using the maximum laser fluence (equal to Δ$T$* = 177 K), the intensity of the martensite peak disappears almost completely within the duration of the laser pulse. Thus, for large overheating the short time of 7 ns is sufficient for the complete martensite to austenite transformation. One question remains, which we address in the following: why does a full transformation require 177 K of heating using a short pulse when around 30 K between austenite start and finish are sufficient in quasistatic experiments?

After the laser pulse, the heat of the thin film is dissipated mostly by heat conduction through the substrate and marginally by radiation loss at the surface. During this cooling process, the sample transforms from austenite to the martensite state. Therefore, the measured intensity increases (Fig. 2(A)). The transformation back to martensite proceeds much slower than the martensite to austenite transformation. The process is slowest for the maximum temperature rise of 177 K, where after the first 250 ns only about half of the sample volume has transformed back into the martensite. A complete backward transition can require up to several hundred µs, when the base temperature lies within the transition region. Indeed, for the maximum laser fluence, the 1 ms spacing of the laser pulses is insufficient for the sample to fully cool down to the original base temperature between the laser pulses. This leads to a slight increase of the base temperature from 328 K to 330 K. Thus, $T_0$ approaches the transformation temperature, which is why for $t < 0$ and $\Delta T^* = 177$ K the intensity is slightly reduced compared to the curves with lower $\Delta T^*$, as evident in Fig. 2(A). It is worth to add that some heat dissipation already occurs during the short heating by the laser pulse [25]. However, our approach to calibrate the temperature rise by diffraction inherently accounts for this. To summarize, the slower dynamics for the transformation during cooling are controlled by heat dissipation, which hinders probing the intrinsic time scale of the austenite to martensite transition. As we aim to understand the fundamental limit, our detailed analysis focuses on the transformation during heating.

In a second series, we varied the base temperature of the sample $T_0$, while using a constant laser fluence equivalent to $\Delta T^* = 177$ K (Fig. 2 (B)). With increasing $T_0$, the sample approaches the austenite finish temperature and accordingly, we already start with a smaller martensite fraction before the laser pulse. When heated with high laser fluence, the sample transforms fast and almost completely for all base temperatures. Afterwards, during cooling, the martensite fraction approaches the initial state again. However, this process is much slower at an increased base temperature. We attribute this to the reduced temperature difference between base temperature and martensite finish temperature, which is driven by the heat dissipation during the end of the transformation. This effect is most pronounced for $T_0 = 354$ K, which according to the arguments in the last paragraph also exhibits the lowest base intensity. This hampered the detailed analysis described later, and thus this measurement was excluded together with the measurement at $T_0 = 345$ K, where we failed to record enough data points for $t < 0$. Nevertheless, the high surface to volume ratio of our thin film leads to an accelerated heat exchange compared to bulk. For a sufficiently low base temperature of 312 K, at least 80 % of the austenite transforms back to the martensite within the first 200 ns – much faster than any report before for a ferroelastic transformation from austenite to martensite.

In this paragraph, we describe how we convert the experimental results to quantitative values that describe how fast one can drive a martensite to austenite transformation. For this, we take the amount of overheating $\Delta T$ (as defined above) as control parameter since this can be directly converted to the driving energy needed for the transformation. In Fig. 3 the driving energy is used as the top axis, which is the added thermal gravimetric energy density obtained by multiplying $\Delta T$ with the mean heat capacity of 500 J kg$^{-1}$ K$^{-1}$ for Ni-Mn-Ga [26,27]. The characteristic values of the transformation, the transition time $\Delta t$ and transformation rate $r$, were obtained by fitting the martensite fraction with a logistic function and are plotted in Fig. 3 for the fluence and temperature series. We define $\Delta t$ as the time interval in which 90 % of the intensity change occurs (see supplementary Fig. S5 for a detailed description). The transformation rate $r$ = (sample fraction transformed)/$\Delta t$ uses the sample fraction transformed, which is the intensity change in relation to the fully martensitic state. The transformation rate exhibits some scatter (cf. Fig. 3), which originates from the low intensity typical for cutting edge time resolved experiments. Both series exhibit the same trend, but the slope of series 2 differs, which we attribute to the reduced intensity in vicinity of the transformation temperature, as discussed in the previous paragraph. This is also the case for the first data point in series 1 at $\Delta T$ = 55 K, where only 30 % of the sample transforms during the laser pulse. Nevertheless, Fig. 3 clearly reveals that increasing overheating reduces the transition time (black curve). The origin of this speed limit is analyzed looking at the red curve, which reveals the increase of transformation rate with driving energy. We used a linear fit on the transformation rate for series 1, which entails two parameters with scientific impact: First, the threshold overheating of about 8 K, which is found by extrapolating the linear fit to $r = 0$, and second, the driving energy accelerating the transformation, which corresponds to the slope of $10^6$ g(Js)$^{-1}$ for the linear fit.

First, we analyze the threshold overheating $\Delta T$ of about 8 K, required for a thermally accelerated transformation rate. Above this threshold, we observe a linear relation between transformation rate and driving energy, which is expected from a general thermodynamic viewpoint, as it represents an Onsager relation [28]. Below this threshold, for martensitic transformations peculiarities are observed, which violate this relation [29–31]. In so-called athermal martensite the transformation only depends on temperature, but not on time. This rate independent behaviour originates from the complex energy landscape exhibing shallow local minima along the transformation path, which result in a discontinuous transformation, proceeding by avalanches during nucleation and growth [32]. Indeed, often an asymmetry between both transformation directions is observed [33,34], and thus microscopic reversibility as one key precondition for an Onsager relation is violated during common, low overheating. However, for the huge overheating used in our experiments, the shallow minima become irrelevant and only the major energy barrier between both phases is important. Accordingly, a martensitic transformation follows a different path when completed at a short time scale. This dynamic reversibility explains why we observe an Onsager relation even at a time scale below 10 ns. Moreover, our observation of an Onsager relation indicates a local thermodynamic equilibrium within this time scale, as this is another prerequisite for this relation. As a transition between athermal and thermally accelerated transformation we propose to use the heating rate, which in our experiments is about $10^9$ K/s at threshold $\Delta T$. For applications, we suggest to remain within the athermal regime, as this avoids the additional energy required to accelerate the transformation. We would like to add, that in our experiments we could only vary the heating rate by laser fluence, but not by pulse duration. Thus, we propose further experiments around the transition region in order to understand the transition between athermal and thermal behavior in more detail.

Second, beyond the athermal regime we measure a slope of the transformation rate of $10^6$ g(Js)$^{-1}$, which describes how much energy is required to accelerate the transformation. To put this value into perspective, we scale it to one $Ni_2MnGa$ unit cell of 242 u, which gives $2.5 \times 10^{27}$ (Js)$^{-1}$. The unit cell is the fundamental entity to distinguish different thermodynamic states of the material. In a crude approximation, a tetragonal martensitic unit cell has three states, as it can point to any of the three possible orthogonal directions. This bit of information is deleted when the sample is heated to cubic austenite. Considering this as a distinct and complete set of thermodynamic states enables a comparison with the Margolus–Levitin theorem [35], which states that the fundamental speed limit for switching one bit of information is $3 \times 10^{33}$ (Js)$^{-1}$. Indeed, this theorem states quite general: 'The average energy $E$ (…) of an isolated physical system tells us directly the maximum number of mutually orthogonal states that the system can pass through per unit of time.' [35] Computation – and martensitic transformations – are thus just two particular examples for this fundamental speed limit.

In this paragraph, we discuss the limitations and accuracy of our experiments. In order to enable these experiments, we had to make a compromise on film thickness. On one side, we need a thick film to maximize the diffracted intensity from our film. On the other hand, the laser light is absorbed just within several tens of nanometers, which results in an inhomogeneous temperature profile over the film thickness. As a compromise, we selected a film thickness of 0.5 µm, with a diffraction efficiency of 75 % of the incident x-rays, and calculated the temperature evolution during and after the laser pulse (supplementary Fig. S3). Taking the values for the maximum laser fluence of 60 mJcm$^{-2}$ as an extreme case, we obtain an average temperature of 600 °C at the end of the laser pulse, whereas the top of the film reaches 820 °C and the bottom is just at 430 °C. This temperature distribution limits the accuracy of our analysis. As the temperature at the surface is about 35 % higher than average, our measurements may overestimate the slope of the transformation rate by this percentage in the pessimistic case that we mostly probe only the hotter surface. On the other side, we also probe the colder, bottom part, which transforms slower. A homogenous temperature profile should therefore result in a faster transformation than our measurements. From the temperature difference between average and bottom we thus expect an underestimation of the transformation time by about 30 %. A further limitation in our experiment originates from the thermal stress due to the temperature difference between the hot film and the cold substrate. As worked out in detail in supplementary section 6, this stress induced martensite can increase the transition temperature between 9 and 53 K when heating up the film by the maximum $\Delta T^* = 177$ K. The large variation originates from the unknown initial stress state and accordingly we cannot correct this contribution to the driving energy of the martensitic transformation. Thus, our measurements underestimate the driving energy by 30 %. To wrap up, probing the speed limit of martensitic transformation is an experimental challenge. Though our measurements of the slope of the transformation rate can have an error up to exceeding 50 %, they give a first idea about the fundamental speed limit of martensitic transformations at all. To improve the experimental accuracy by future synchrotrons with higher brilliance, we propose to use thinner films as they exhibit a more homogenous temperature profile. To eliminate the stress between film and substrate, freestanding films might be used, but their handling is difficult when they are thin.

Though our integral measurements only reveal how fast the transformation of a whole sample can occur, for future experiments it is interesting to consider also the underlying microstructural processes, which at common time scales consist of nucleation and growth, as described within the introduction. Coexistence of both phases is a prerequisite for both processes, and our experiments indicate that this is also the case at this time scale (Fig. 1). However, the observed coexistence may also originate from the inhomogeneous temperature profile, as discussed in the last paragraph. Furthermore, the temperature inhomogeneity can have an impact on the nucleation process. In particular, some martensite may have remained in the colder part of the film, which can act as nuclei and thus accelerate the transition. At the huge overheating used in our experiments, however, nucleation may not be the limiting process anyway, since even the classical homogenous nucleation model predicts an exponential increase of nuclei. In case that many nuclei are present, they do not need a high growth velocity to transform the whole sample. An SEM micrograph of the present film (Supplementary Figure S6) reveals that already after slow cooling a finely twinned microstructure is present. More fundamentally, our observation of a thermally accelerated transformation in contrast to athermal transformations at common time scales, indicate for a fundamentally different transformation path, as discussed above. Accordingly, we expect substantial changes in the martensitic microstructure after fast cooling. We propose to use dedicated methods with a high spatial and temporal resolution [17] to probe microstructural changes in detail. We speculate, that the decisive microstructure consists of some $10^6$ unit cells, which could explain the difference between the speed limit we observe and the ultimate speed limit by the Margolus–Levitin theorem, which assumes that each unit cell can switch independently.

## 4. Conclusions

To conclude, our dedicated setup allows probing the speed limit of a martensite to austenite transformation. This transformation can be completed below 10 ns, but requires an overheating of several hundreds of Kelvin. For most applications, only a much smaller overheating can be reasonably applied, but also then the switching time is still in the sub-microsecond scale. The austenite to martensite transformation, although limited by heat dissipation in our experiments, can be completed at least within hundreds of ns. This leaves plenty of room to increase the performance of all applications based on martensitic transformations. Our analysis further reveals that driving a martensite to austenite transformation is limited by the product of energy and switching time of around $2.5 \times 10^{27}$ $(Js)^{-1}$, which approaches the ultimate limit determined by the Margolus–Levitin theorem. Thus, this speed limit has the same origin as in microelectronics, where today's computers just reach $10^{10}$ $(Js)^{-1}$ – a value which increased exponentially over the years following Koomey's law [36].

Despite the fundamental speed limit, the overall performance of microelectronic devices still increases due to the ongoing miniaturization. This path is also accessible for martensitic microsystems since the required total energy is proportional to the volume. Furthermore, the reduced size accelerates heat dissipation, which is decisive for a fast martensite to austenite transformation. Thus, the ultimate performance of martensitic microsystems may just be limited by finite size effects – a topic not yet examined at all.

Acknowledgements

Parts of this research were carried out at Petra III at DESY, a member of the Helmholtz Association (HGF). The authors would like to thank Dimitri V. Novikov for assistance in using beamline P23. The authors acknowledge Roman Bauer for the support during the measurements at DESY as well as Anett Diestel for the sample preparation and helpful discussions. We thank Tilmann Hickel, Matthias Bickermann, and Jürgen Lindner for helpful discussions.

Funding:

This work was supported by the German Research Foundation under grant GA 2558/5-1 (PG) and the German Research Foundation under grant FA 453/11 via priority program SPP 1599 (SF).

Disclosure statement:

The authors report there are no competing interests to declare.

Data and materials availability:

Measured synchrotron data and calculated thermal evaluation during irradiation with the laser pulse is available at https://rodare.hzdr.de/ DOI: 10.14278/rodare.1465.

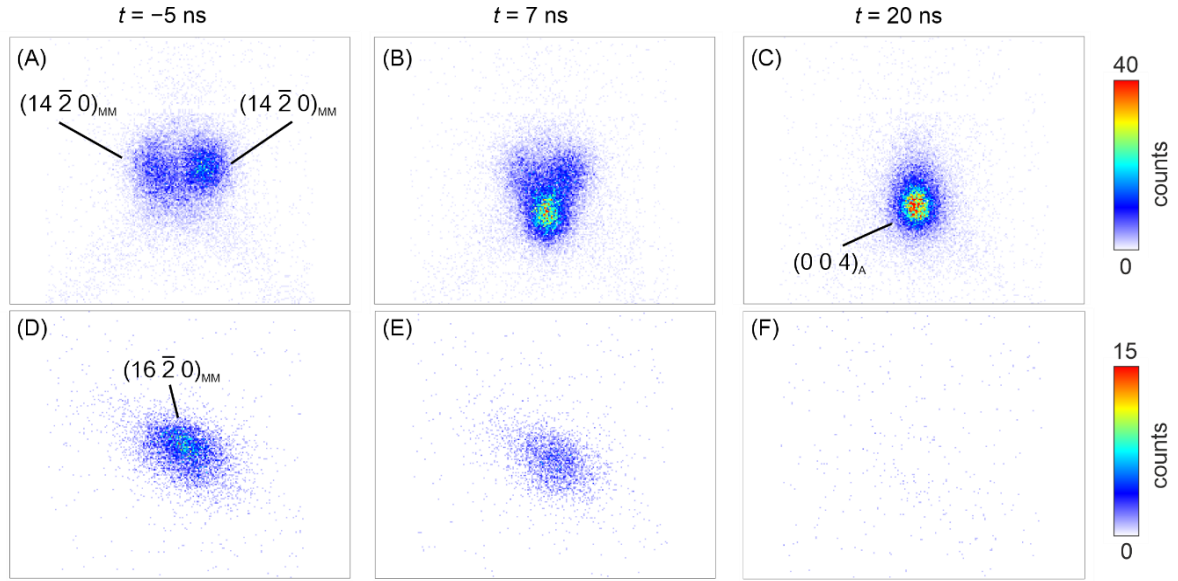

Fig. 1. Detector images taken while heating with a laser pulse reveal the time-dependent transformation from martensite to austenite. The images in the first column are taken in the martensite state, 5 ns before the pulse (A, D); those in the second column at the end of the pulse at 7 ns, where both phases coexist (B, E); and those in the third column after the more or less complete transformation of the sample region at t = 20 ns (C, F). Both rows show the raw data in detector pixel coordinates. The pictures in the first row (A-C) show the area around the $(0\,0\,4)_A$ and neighboring $(14\,\bar{2}\,0)_{MM}$ reflections; the second row shows the area around $(16\,\bar{2}\,0)_{MM}$, where no overlap with the austenite occurs (D-F). These exemplary diffractograms have been taken at an initial sample temperature of $T_0 = 312$ K while heating with a laser pulse of 60 mJ cm$^{-2}$, which results in a temperature increase of $\Delta T^* = 177$ K. A full time series of a larger region around $(0\,0\,4)_A$ is available as a supplementary video. Indexing follows [20] where the index A denotes an austenite reflection and MM one of modulated martensite. The intensity in all images is scaled linearly according to the scale given on the right side for both rows.

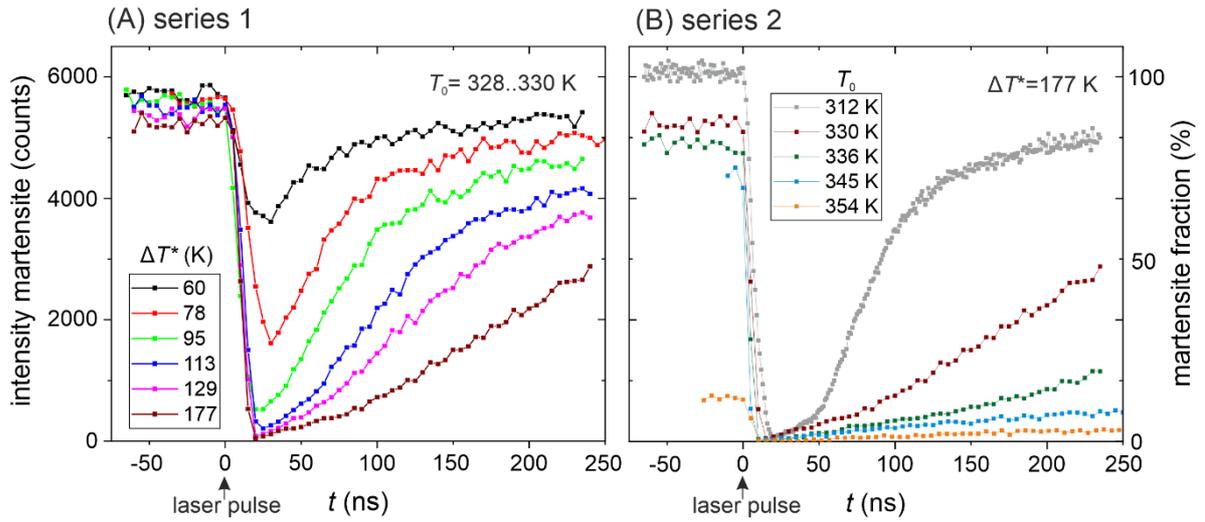

Fig. 2. Probing the time dependency of a martensitic transformation while heating with a laser pulse and subsequent cooling. The summed up intensity of the (16 $\bar{2}$ 0)$_{MM}$ peak allows to determine the martensite fraction, which reduces sharply during heating with the 7 ns laser pulse and increases afterwards when the sample cools down. Two series with different experimental conditions were investigated: (A) Nearly the same base temperature of around $T_0$ = 329 K was used for all measurements and the laser fluence was varied to obtain temperature rises ΔT* between 60 K and 177 K. (B) The laser fluence and therefore ΔT* was kept constant at 177 K, but the base temperature was varied between 312 K and 354 K.

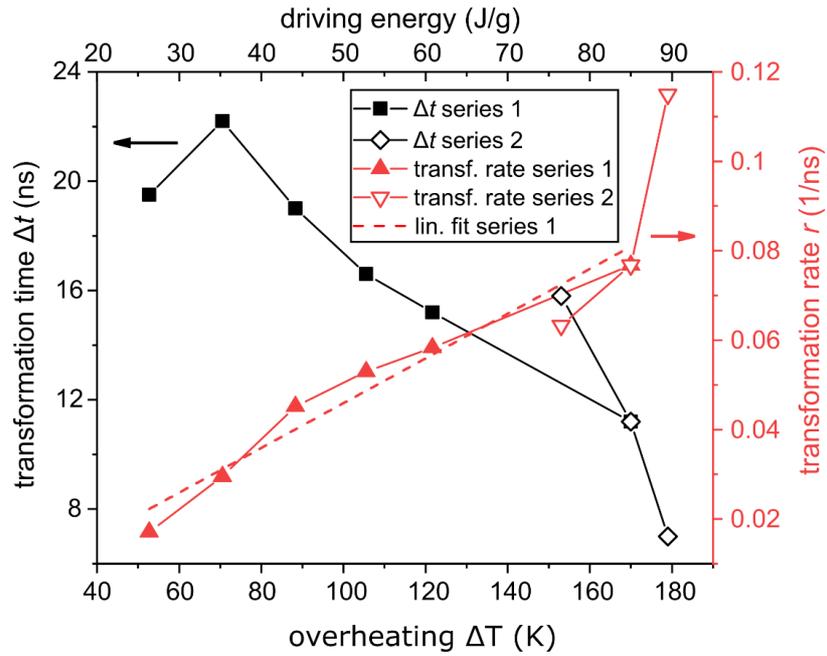

Fig. 3. Driving a martensite to austenite transformation as fast as possible. To complete the transformation in a short time interval Δ*t*, a sufficient overheating Δ*T* above the austenite start temperature by the 7 ns laser pulse is required (bottom axis). The laser pulse drives the transformation by adding thermal energy to the sample, which is plotted as additional top axis. An increase of driving energy increases the transformation rate *r* (right axis). This graph contains data obtained for series 1 (solid symbols) and series 2 (open symbols); the latter exhibits a larger experimental error, as described in the text. Therefore, we use only data from series 1 for the linear fit.

# Supporting Information

**What is the speed limit of martensitic transformations?**

*Stefan Schwabe, Klara Lünser, Daniel Schmid, Kornelius Nielsch, Peter Gaal, Sebastian Fähler\**

*\* s.faehler@hzdr.de*

**1) Set-up for pulsed laser heating and time resolved synchrotron diffraction**

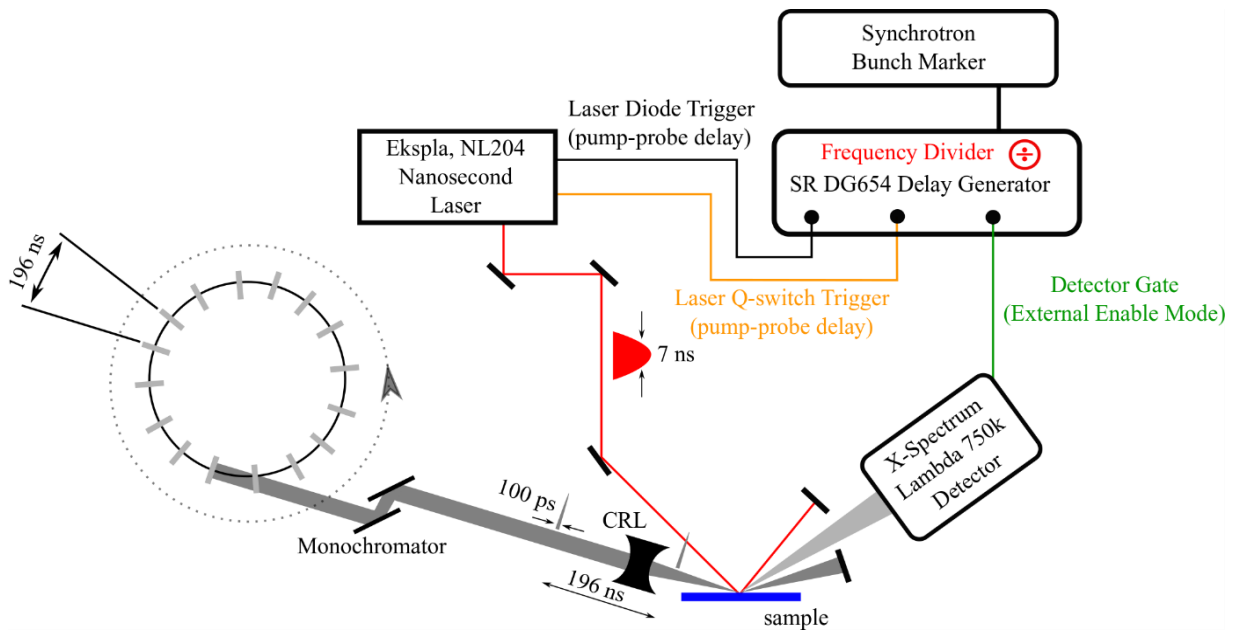

Fig. S1: **Sketch of the experimental setup.** A q-switched laser (Ekspla, NL204) is synchronized to the Petra III bunch marker using a Stanford Research DG654 Delay Generator and delivers optical pulses at a wavelength of 1064 nm with a pulse energy and a pulse duration of up to 2 mJ and 7 ns, respectively. 100 ps X-ray pulses from the Petra III storage ring are monochromatized and focused onto the sample to probe transient structural dynamics in the sample. The diffracted signal is captured by a Lambda 750k area detector in external gating mode. The gating reduces the acquisition rate to 1 kHz, i.e., the same rate as the optical excitation pulses. This sketch was adapted from [37].

**2) Estimated radiation transmission rate**

To estimate if the synchrotron radiation probes the complete film thickness, we calculated the ratio of transmitted X-ray for our film.

$$\frac{I_1}{I_0} = e^{-\mu x}$$

$I_0$ and $I_1$ are the X-ray intensities before and after the film, $\mu$ is the linear attenuation coefficient of the material and $x$ the distance travelled in the film. To give information about the whole film, the radiation has to travel through the film to the substrate and back out again. As the radiation hits the film in an angle of 18° and with a film thickness of 500 nm, $x$ equals 3200 nm. With $\mu$ of 887 cm$^{-1}$ (for Ni$_2$MnGa with a density of 8 g cm$^{-3}$ and a beam energy of 12.1 keV, [38]), $\frac{I_1}{I_0}$ calculates to 75 %, which means that still a reasonable amount of the x-ray radiation is not absorbed in the film. Therefore, we can assume that the synchrotron radiation gives information from the complete film thickness.

**3) Characterization of the temperature dependent phase transition**

To select reasonable base temperatures for our pump-probe experiments, a thorough understanding of the transition temperatures of the material is necessary. Therefore, the phase transition of the investigated Ni-Mn-Ga thin film was studied using magnetization as well as resistivity measurements, in addition to the diffraction experiments. Fig. S2 (a) shows the temperature-dependent magnetization of the sample, which was measured by a vibrating sample magnetometer (VSM) in a PPMS system (VersaLab$^{TM}$, 2 K/min heating and cooling rate in an external magnetic field of 0.01 T). The phase transition shows a hysteresis and is accompanied by a change in magnetization at around 350 K. As the transition partly overlaps with the Curie-temperature at 366 K, an additional resistivity measurement was performed in another PPMS system. The resulting curve in zero field is shown in Fig. S2 (b) in black and for clarity only the reverse transition is plotted. As the austenite has a lower resistivity than the martensite, there is a clear drop at the phase transition. For comparison, the phase fraction of the martensite was also measured at the beamline using the diffracted intensities of the (16 -2 0)$_{MM}$ martensite peak without laser heating. The red curve in Fig. S2 (b) shows this phase fraction for particular temperatures, which were held steady within ±0.3 K for the measurement time. To obtain the martensite phase fraction, the intensity was normalized to the maximum intensity at 306 K. Both curves show a similar transition region, which is shifted to slightly lower temperatures for the curve derived from the diffraction experiments. This may be attributed to the different measurement setups, as the thermometers are placed at different distances to the sample. Furthermore, during the resistivity measurement, the temperature was swept with a rate of 3 K/min compared to the quasi-static diffraction experiments. To determine the austenite start and finish temperatures, we thus used the curve obtained by X-ray diffraction as the corresponding temperature measurement setup was the same one used for all time-dependent measurements at the synchrotron. As with the time dependent measurements, we defined the austenite start temperature as the point were 5 % of the sample has transformed and the finish temperature for 95 % transformation, which gives 336 K and 365 K, respectively. An interesting point that can only be seen from the diffraction experiment (red curve) is that even at 400 K – considerably above the sharp intensity drop – the transition to the austenite is not fully completed yet. This aspect is only visible in the direct diffraction experiments and not by the commonly used indirect methods like magnetization and resistivity.

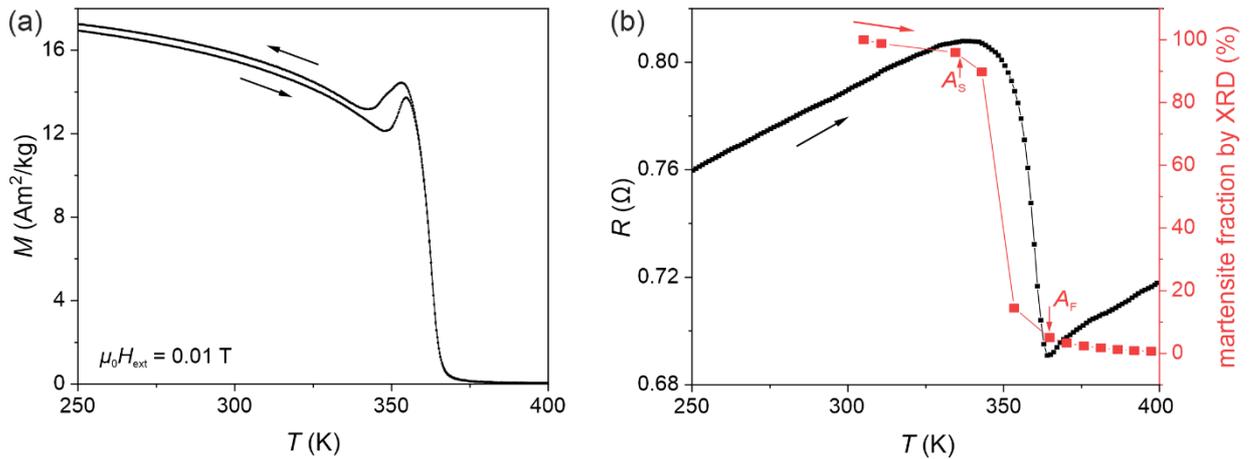

Fig. S2: **Determining transition temperatures in quasi-static measurements.** (a) Temperature dependent magnetization of the investigated sample measured with VSM in a PPMS system (VersaLab™). An external field of 0.01 T was applied. As the phase transition partially overlaps with the Curie-temperature ($T_c$ = 366 K), an additional resistivity measurement was performed in another PPMS system in zero field ((b), black curve). For comparison, the temperature-dependent intensity of the martensite phase measured quasi-statically at the beamline is shown in red in vicinity of the transformation temperature. Both curves in (b) only depict the reverse transformation. All measurements reveal a similar transformation behavior and slight differences in temperature can be attributed to the different setup and sweep rates. Therefore, we used the values obtained directly at the beamline: austenite start $A_S$= 336 K at 95 % martensite fraction and austenite finish $A_F$= 365 K at 5 %, as marked in the graph.

**4) Determination of the laser induced temperature rise**

For the detailed evaluation of the results gathered from the time-dependent diffraction experiments, it is necessary to probe the temperature rise $\Delta T$ induced by the laser. To obtain this, we use the high accuracy of diffraction to probe lattice parameters, which allows to measure the thermal expansion. Thus, the lattice constant of the Ni-Mn-Ga film itself is the "thermometer", and accordingly $\Delta T$ is representative of the relevant sample region and thickness. The calculation of the X-ray radiation transmitted for the present film (section 2) of the supplementary) reveals a diffraction efficiency as high as 75 %, and accordingly we consider these measurements a good average over the film thickness. For these measurements we heated the sample to 413 K into the fully austenite state to avoid any possible variant reorientation within the martensitic state. We chose the maximum laser fluence of around 60 mJ cm$^{-2}$ and monitored the position of the austenite (004) peak perpendicular to the substrate. From that, we derived the change in lattice parameter ($\Delta a_0$) in relation to the initial value before the laser pulse hit the sample, which is plotted in Fig. S3.

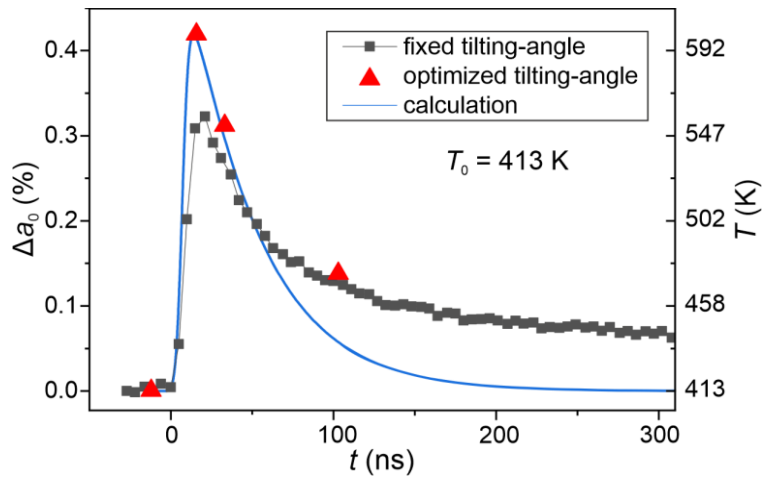

Figure S3: **Determining the laser induced temperature rise.** Time-dependent change of lattice parameter of the austenite ($a_0$) measured while heating the sample with a laser fluence of 60 mJ cm$^{-2}$ from a starting temperature of 413 K. Respective temperatures obtained from a calibration at quasistatic temperatures are shown on the right y-axis. The gray data points were collected with constant diffractometer tilt angles. The red data points show some discrete values with diffractometer angles adjusted for the maximum intensity of the (004)$_A$ austenite peak at selected points in time, as described in detail within the text. The blue line shows the calculated temperature development.

The lattice parameter of the film increases sharply while the sample is heated up and decreases afterwards while the sample "slowly" cools down. The black points were recorded continuously with a constant set of diffractometer angles. However, the fixed detector and change of lattice parameter can mean that one does not perfectly hit the austenite reflection in reciprocal space, which may result in an apparent slight variation in lattice parameter. To exclude this possible error, additional experiments where performed, where for some selected times the diffractometer angles were optimized for intensity. These measurements are plotted in red and show some deviations from the fixed diffractometer measurements, especially in the time interval directly after the laser pulse hit the sample. Therefore, the red data point with the maximal change of lattice parameter is used further on.

To calibrate our "Ni-Mn-Ga thermometer", we measured the thermal expansion coefficient in out of plane direction for our sample in a laboratory diffractometer and obtained a value of $22.4 \times 10^{-6}$ K$^{-1}$. Though we used the same sample for calibration, the accuracy of this approach is limited by the different temperature profile and thermal expansion (see section 7 of this supplementary), since during the calibration the complete substrate is hot, whereas during pump probe experiments most of the thick substrate remains at ambient (see section 5 of the supplementary). This calibration allows us to convert the maximum change in lattice parameter into a temperature change, which results in $\Delta T^* = 187$ K for a laser fluence of 60 mJ cm$^{-2}$. The conversion for all the other laser fluence values used in our investigation was done assuming a linear relationship between laser fluence and temperature change. To account for the latent heat of the transformation, we compared the latent heat to the specific heat capacity of Ni-Mn-Ga. With a latent heat of around 5000 J kg$^{-1}$ [39,40], and a specific heat capacity of around 500 J kg$^{-1}$ K$^{-1}$ [26,27], around 10 K of the $\Delta T$ will be absorbed by the material as latent heat. We compared the measured values with calculations using the thermal and optical properties of Ni-Mn-Ga (blue line in Figure S3). The maximum temperature change of $\Delta T^* = 187$ K is confirmed by the calculation, described in detail in the next section. However, according to the calculation, the temperature decreases faster after the laser pulse than in the experiment. We attribute this to the film-substrate interface that acts as a heat transfer coefficient, which is not taken into account in the calculation.

**5) Calculation of temperature profile during and after laser pulse**

To calculate the depth dependent temperature of film and substrate we use the python-based toolbox *udkm1Dsim* [41,42]. This toolbox uses a simplified molecular dynamics model to calculate one-dimensional propagation of coherent and incoherent acoustic phonons. A number of previous measurements showed that at these short laser pulses and large spot sizes heating and subsequent cooling occurs exclusively in the direction perpendicular to the sample surface, i.e., a one-dimensional modelling is sufficient [23]. The simulated structure consists of 1000 unit cells of Ni-Mn-Ga (585 nm) and 30000 unit cells of MgO substrate (12.63 µm). The large thickness of the substrate allows employing a constant temperature boundary condition at the rear side of the substrate. At the sample surface we employ a thermal isolation boundary condition. The complex index of refraction for $Ni_2MnGa$ was taken from literature [43] and slightly adapted to match the observed peak temperature rise in the Ni-Mn-Ga film after optical excitation. The value employed in the simulations was 2.0+1.32j. Thermal properties of bulk Ni-Mn-Ga and MgO were taken from [43] and [44], respectively. The optical excitation pulse is absorbed in the upper 65 nm Ni-Mn-Ga film. At this pulse duration, heat within the film dissipates already during absorption of the laser pulse into the substrate. Thus, the generated peak temperature is approximately 10 times lower compared to the excitation with a ultrashort (sub 1 ps) optical pulse. Figure S4 (a) displays the temperature as a function of time and film depth obtained from our simulations. The temperature at the middle of the film matches well the mean film temperature (see Figure S4 (b)).

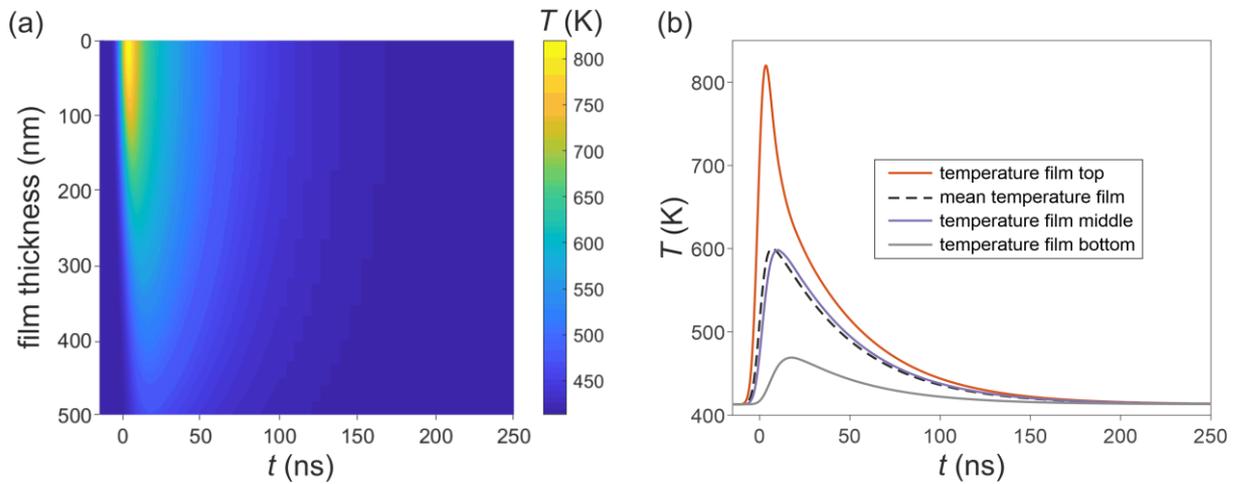

Fig. S4: **Calculated temperature profile of the Ni-Mn-Ga film on MgO substrate during and after the laser pulse**. The temperature as a function of the film depth and the time was calculated with finite elements. The calculations assumed a start temperature of 413 K and a laser fluence of 60 mJ cm$^{-2}$ (a) The heat map shows the evolution of the temperature throughout the film after the laser pulse at t = 0 ns. (b) Temperature profile at specific depths of the film (top, middle and bottom side). The dashed curve displays the mean temperature of the complete film.

**6) Obtaining characteristic values from the time-dependent intensity measurements**

As described in the main part of the paper, our thin film sample was investigated for various different laser fluences as well as base temperatures. To compare the time-dependent phase transition for these different experimental conditions, we first had to extract some characteristic parameters from the intensity data. Starting with the impact of the laser pulse, the intensity curves for the martensite to austenite transition are Z-shaped as more and more of the sample transforms until a maximum of transformed phase is reached. To derive characteristic values from these curves, we fitted the part of the curve in which the intensity decreases using a generalized logistic function. This function can be used to describe saturation and transition processes and requires only a small number of fitting parameters. Three exemplary curves with the corresponding fit (red) are shown in Figure S5 (a-c). The generalized logistic function to describe the time dependency of intensity $I(t)$ has theform:

$$I(t) = \text{UB} - \frac{\text{UB} - \text{LB}}{(1 + e^{-b \cdot (t - t_0)})^{\frac{1}{v}}}$$

UB refers to the upper boundary of the function, i.e., the maximum it approaches. LB is the lower boundary, $t_0$ shifts the position along the time axis, and the parameters $b$ and $v$ describe the shape of the drop. $b$ is the transition rate and $v$ affects how the asymptote is approached. After fitting this function to our data, we derived the following key parameters of a transition process: while UB and LB can be taken directly from the fit, we are additionally interested in the transition time $\Delta t$, which is sketched in blue in Figure S5 (a). We define $\Delta t$ as time span required for the fit function to decrease from 95 % to 5 % of its maximum drop (UB − LB). This approach gives the transition time, which is the key parameter to understand the speed of a martensitic transformation. In particular, this approach allows neglecting details of the curve shape, which may be affected e, g. by the inhomogeneous temperature profile, described in supplementary section 6). We use the generalized logistic function, as it directly give UB and LB and a slightly better fit to our data compared to the standard one. Together with $\Delta t$ these data could also be extracted directly from our measurements by the classical tangent method, but with lower accuracy. For completeness, all fit parameters are summarized in the following table.

Table S1: Fit parameters of the generalized logistic function on Series 1 and 2. $\Delta T^*$ and $T_0$ in the first two rows give the parameters used for the respective measurements. For $\Delta T^*$ of 177 K and $T_0$ of 336 K, we did not measure enough data points to make a fit possible.

| $\Delta T^*$ (K) | 60   | 78   | 95   | 113  | 129  | 177  | 177  | 177 | 177  | 177 |
|------------------|------|------|------|------|------|------|------|-----|------|-----|
| $T_0$ (K)        | 330  | 330  | 330  | 330  | 330  | 330  | 312  | 336 | 345  | 354 |
| UB               | 5735 | 5618 | 5604 | 5502 | 5390 | 5230 | 6081 | -   | 4893 | 717 |
| LB               | 3711 | 1629 | 383  | 156  | 0    | 0    | 0    | -   | 0    | 0   |
| $t_0$            | 6.3  | 7.9  | 2.8  | 9.9  | 4.5  | 4.8  | 9.6  | -   | 3.5  | 6.9 |
| $b$              | 0.3  | 0.2  | 0.2  | 0.3  | 0.3  | 0.4  | 0.2  | -   | 0.9  | 0.8 |
| $v$              | 0.7  | 0.3  | 0.3  | 0.7  | 0.2  | 0.2  | 0.02 | -   | 1.2  | 1.4 |
| $\Delta t$ (ns)  | 20   | 22   | 19   | 17   | 15   | 11   | 16   |     | 7    | 9   |

A generalized logistic function can describe the data points quite well. For the measurement shown in (a), we used a small time increment describing the intensity drop very detailed. Due to the limited measurement time at a synchrotron, we had to increase the step size for most of the experiments as can be seen in (b) and (c). The fit is still able to reproduce the data, which is also the case when the sample only partially transforms as shown in Figure S5 (c). Depending on the experimental conditions, the transition time differs noticeable from well below 10 ns in (b) to nearly 20 ns in (c).

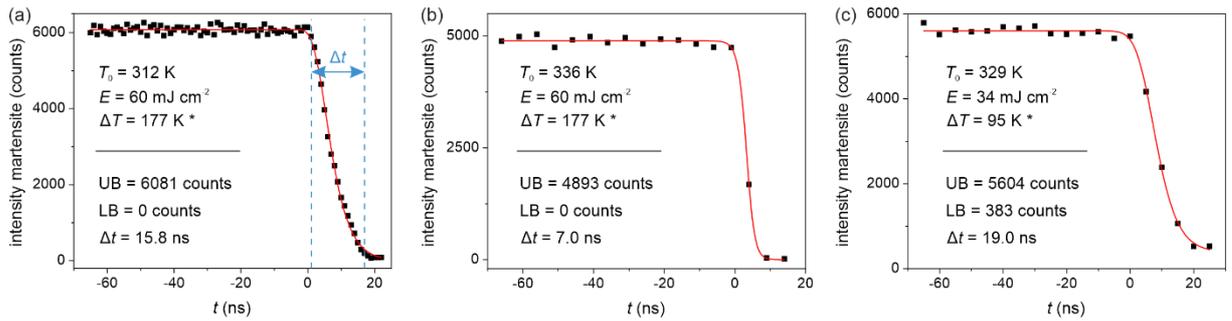

Fig. S5: Analyzing time-dependent measurements while heating the sample with a nanosecond laser pulse (exemplary measurements). The intensity of the martensite $(16\ \text{-}2\ 0)_{MM}$ peak (black) was fitted with a generalized logistic function (red), as described in the text. The upper (UB) as well as lower boundary (LB) of the function and the derived transition times ($\Delta t$, sketched in blue) are given in the figure. The base temperature before the laser pulse ($T_0$) was 312 K (a), 336 K (b) and 329 K (c). The laser fluence (E) and the temperature rise $\Delta T$ is given in the corresponding graphs as well. As described in supplementary section 4), $\Delta T$ was corrected for the latent heat during the transition (*) by subtracting 10 K from the values obtained from the measurements shown in Fig. S3.

**7) Estimation of film stress and influence on transformation temperature**

In our analysis we focus on the influence of temperature, but the difference of thermal expansion of the hot, thin film on the cold, thick substrate can result in a film stress, which commonly increases the martensitic transformation temperature. Following our analysis of thermal stress in Ni-Mn-Ga films [45] we estimate the increase of transition temperature by thermal stress for the maximum temperature rise of 177 K. From the thermal expansion coefficient of 22.4 x $10^{-6}$ $K^{-1}$ of Ni-Mn-Ga (see section 4) of this supplementary) we expect a strain of 0,4 %, which is equivalent to a stress of 120 MPa when using the E-module = 20 GPa of austenite [46]. Following the Clausius-Claperyon equation of Ni-Mn-Ga [47,48], this stress increases the transition temperature by 54 K. This, however, is an upper estimate, as this stress is compressive, and films after growth often exhibit tensile stress [45], which can compensate most of the thermal stress. When using the tensile stress of this paper as an approximation (the stress of the present film is not known), the increase of transition temperature is reduced to 9 K. When putting these values in relation to the temperature rise of 177 K, we underestimate the driving energy by 5 % to 30 % by neglecting thermal stress.

We would like to add, that at no time the surface temperature of our film (Fig. S4a) approaches the melting point and thus no shock wave by evaporating forms, as used in other dedicated setups [49].

## 8) Microstructure of the examined Ni-Mn-Ga film

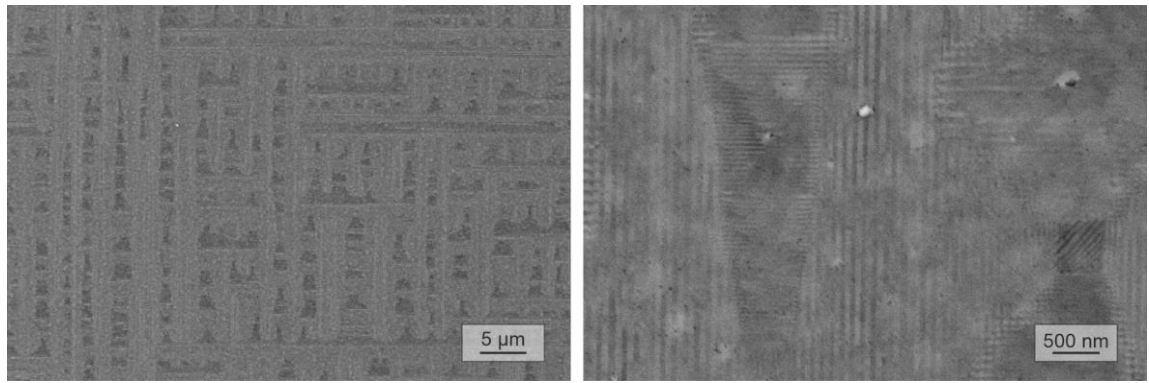

Fig. S6: SEM micrographs of the examined Ni-Mn-Ga film in different magnification. The martensitic microstructure consists mostly of Type Y and the mesoscopic twin boundaries are 20…50 nm apart. We would like to add that we recently published a comprehensive explanation how and why this hierarchical microstructure forms (after slow cooling) [22].